# Scaling and memory in seismological phenomena*

Sumiyoshi Abe[1,2,3] · Norikazu Suzuki[4]

[1] Department of Physics, College of Information Science and Engineering,
  Huaqiao University, Xiamen 361021, China
[2] Institute of Physics, Kazan Federal University, Kazan 420008, Russia
[3] Department of Natural and Mathematical Sciences,
  Turin Polytechnic University in Tashkent, Tashkent 100095, Uzbekistan
[4] College of Science and Technology, Nihon University, Chiba 274-8501, Japan

**Abstract**

The concept of memory is of central importance for characterizing complex systems and phenomena. Presence of long-term memories indicates how their dynamics can be less sensitive to initial conditions compared to the chaotic cases. On the other hand, it is empirically known that the Feller-Pareto distribution, which decays as the power law i.e. the scale-invariant nature, frequently appears as a statistical law generated by the dynamics of complex systems. However, it is generally not a simple task to determine if a system obeying such a power law possesses a high degree of complexity with a long-term memory. Here, a new method is proposed for characterization of memory. In particular, a scaling relation to be satisfied by any memoryless dynamics generating the Feller-Pareto power-law distribution is presented. Then, the method is applied to the real data of energies released by a series of earthquakes and acceleration of ground motion due to a strong earthquake. It is shown in this way that the sequence of the released energy in seismicity is memoryless in the event time, whereas that of acceleration is memoryful in the sampling time.

**Keywords**    New method for characterization of memory · Scale invariance ·
          Feller-Pareto distribution · Earthquake energy · Acceleration of
          ground motion

# Introduction

Dynamics governing a complex system contains long-range correlation and long-term memory, that is, spatio-temporal nonseparability, in general.

Consider an ordered sequence i.e. a process $\{x_1, x_2, ..., x_\mathcal{N}\} \equiv \{x_n\}_{n=1, 2, ..., \mathcal{N}}$, empirically obtained by the measurements of a certain physical quantity of a given system and suppose that the map from $x_n$ to $x_{n+1}$ in the sequence as a deterministic/nondeterministic dynamics (Schuster and Just 2005) is largely unknown. What is frequently performed is to investigate the "statistical nature" of the sequence. A form of a resulting distribution is, however, not sufficient for extracting information about the unknown dynamics. Then, the next step is to clarify the property of memory. (Clearly, only temporal correlation is relevant to one-dimensional dynamics.)

Problems concerning long-term memory are ubiquitous in contemporary science. They are of particular importance if a system exhibits non-Gaussian statistics (Riggs and Lalonde 2017). To examine if memory is contained in a sequence, the simplest way may be to examine random shuffling of the order of events. A method better than it, which is widely used, is to analyze the autocorrelation function (Kantz H, Schreiber T 2004). It is however nontrivial to quantify the memory from such a function, in general, because of the fluctuation of the function due to finite data size, for example. This point is particularly salient if a distribution is of the power-law type. Thus, any novel attempt on this issue should be welcome, and therefore we wish to present one such, here.

Today, it is in fact well noticed that there are abundant systems, in which distributions of physical quantities are not of the familiar exponential or Gaussian types but of the power law. The Gutenberg-Richter law serves as a typical example. Thus, in this paper, we address ourselves to development of a new method for characterizing memories in systems obeying the Feller-Pareto (asymptotic) power-law distribution that decays as a power law (Clark, Cox and Laslett 1999). In particular, we present a new scaling relation to be satisfied by any memoryless dynamics generating the Feller-Pareto distribution. Then, we apply it to the real data of seismicity: released energy and acceleration of ground motion. We show that this method reveals in a peculiar way that the sequence of the released energy is memoryless, whereas the sequence of the acceleration of ground motion is memoryful.

## New method and scaling relation for memoryless dynamics generating the Feller-Pareto power-law distribution

Let us take from the total ordered sequence $\{x_n\}_{n=1, 2, ..., \mathcal{N}}$ its two ordered subsequences: $\{x_k\}_{k=1, 2, ..., N-m}$ and $\{x_{k+m}\}_{k+m=m+1, m+2, ..., N}$, where $1 << N \leq \mathcal{N}$ and $m$ is the shift index, $m = 1, 2, ..., K$, with $K < N$. Let the joint distribution associated with these two subsequences be $F(x, x')$, where $x$ and $x'$ denote the first and second ones, respectively. Our starting point is to reduce it to a single-variable distribution. To do so, we consider a certain bivariate quantity, $Q(x, x')$, and see how it distributes. Such a distribution is given by $P(r) = \langle \delta(r - Q) \rangle$, where the angle brackets stand for the average with respect to the joint distribution:

$$P(r) = \iint dx\,dx'\,\delta\bigl(r - Q(x, x')\bigr) F(x, x'), \tag{1}$$

where the continuous notation is used. Then, our idea is based on the fact that the process is strictly memoryless, if the joint distribution is factorized: $F(x, x') = f(x) g(x')$. Moreover, in the case when the sequence is *stationary* for sufficiently large $N$, $g(x')$ is identical to $f(x')$. Therefore, in such a case, the joint distribution has the form

$$F(x, x') = f(x) f(x'), \tag{2}$$

which is a condition to be satisfied if the sequence is stationary and memoryless. It is noted that parameters contained in the function *f* may depend on the shift index *m*, in general.

Now, we are interested in a class of complex systems obeying the normalized distribution,

$$f(x) = (\alpha - 1) c^{\alpha - 1} \times \frac{1}{(c + x)^{\alpha}} \qquad (\alpha > 1), \tag{3}$$

defined in the half space $x \in (0, \infty)$, where *c* is a positive constant. This is a shifted power-law distribution that is free from the singularity in the limit $x \to 0+$. Although it actually belongs to a more general Feller-Pareto family containing 5 parameters (Clark, Cox and Laslett 1999), here and hereafter it is simply referred to as the Feller-Pareto distribution.

Since the distribution in Eq. (3) decays as

$$f(x) \sim \frac{1}{x^\alpha}, \tag{4}$$

it is natural to impose the scale invariance also on $Q(x, x')$, that is,

$$Q(\lambda x, \lambda x') = Q(x, x'), \tag{5}$$

where $\lambda$ is an arbitrary positive constant. Our choice here, which may be simplest, is the following one:

$$Q(x, x') = \frac{x'}{x}. \tag{6}$$

In this case, $P(r)$ in Eq. (1) with Eq. (2) is expected to also decay as a power law

$$P(r) \sim \frac{1}{r^\beta} \quad (\beta > 1). \tag{7}$$

The exponents in Eqs. (4) and (7) are not independent each other and should satisfy a certain relation. Below, we show that in fact the relation,

$$\beta = \begin{cases} \alpha & (1 < \alpha < 2) \\ 2 & (\alpha > 2) \end{cases}, \tag{8}$$

which may be regarded as a scaling relation, holds for a memoryless stationary sequence generating the Feller-Pareto distribution.

In the context of Markovian processes, the scaling relation for the number of events and the waiting-time distribution, both of which are asymptotically scale-invariant, has been derived (Barndorff-Nielsen, Benth and Jensen 2000). Originally, it has been studied for the problems of subrecoil laser cooling of atoms (Bardou, Bouchaud, Aspect and Cohen-Tannoudji 2002) and later has successfully been applied to earthquake aftershocks (Abe and Suzuki 2009, 2012) and granular materials (Tsuji and Katsuragi 2015) to examine if the processes are (non-)Markovian. In particular, the process of earthquake aftershocks has been found to be non-Markovian (Abe and Suzuki 2009, 2012).

To prove Eq. (8), first we substitute Eqs. (2), (3) and (6) into Eq. (1) to have

$$P(r) = (\alpha-1)^2 c^{2\alpha-2} \int_0^\infty dx \frac{x}{(c+x)^\alpha (c+rx)^\alpha}$$

$$= \frac{(\alpha-1)^2}{r} \int_0^\infty dx \frac{x}{(1+2zx+x^2)^\alpha} \qquad (9)$$

with

$$z = \frac{1}{2}\left(\sqrt{r} + \frac{1}{\sqrt{r}}\right), \qquad (10)$$

provided that we have used $\delta(r - x'/x) = x\delta(rx - x')$ and the change of the integration variable $x \to cx/\sqrt{r}$ to obtain the first and second equalities, respectively. It should be noticed that $P(r)$ in Eq. (9) does not depend on the constant $c$. This fact has its origin in the manifest scale invariance in Eq. (5) (see the discussions in the end of this section as well as the final section). The integral on the right-hand side in Eq. (9) is calculated as follows:

$$\int_0^\infty dx \frac{x}{(1+2zx+x^2)^\alpha} = 2^{\alpha-1/2}(z^2-1)^{1/4-\alpha/2}\,\Gamma\!\left(\alpha+\frac{1}{2}\right) B(2\alpha-2,2)\, P_{\alpha-5/2}^{1/2-\alpha}(z), \qquad (11)$$

where $\Gamma(s)$, $B(p,q)\,[=\Gamma(p)\Gamma(q)/\Gamma(p+q)]$ and $P_\nu^\mu(z)$ are the Euler gamma function, the beta function and the associated Legendre function, respectively (Gradshteyn and Ryzhik 1980, formula 3.252.11). Therefore, we have

$$P(r) = (\alpha-1)2^{2\alpha-3}\,\Gamma\!\left(\alpha-\frac{1}{2}\right)\frac{1}{r}\left|r-\frac{1}{r}\right|^{1/2-\alpha} P_{\alpha-5/2}^{1/2-\alpha}(z) \qquad (12)$$

with $z$ in Eq. (10). Finally, using the asymptotic form of the associated Legendre function

$$P_{\alpha-5/2}^{1/2-\alpha}(z) \sim \frac{1}{\sqrt{\pi}}\begin{cases} \Gamma(2-\alpha)\,(2z)^{3/2-\alpha} & (1<\alpha<2) \\[2mm] \dfrac{\Gamma(\alpha-2)}{\Gamma(2\alpha-2)}(2z)^{\alpha-5/2} & (\alpha>2) \end{cases}, \qquad (13)$$

we obtain

$$P(r) \sim \frac{\alpha-1}{\sqrt{\pi}} 2^{2\alpha-3} \Gamma\left(\alpha-\frac{1}{2}\right) \times \begin{cases} \dfrac{\Gamma(2-\alpha)}{r^{\alpha}} & (1 < \alpha < 2) \\ \dfrac{\Gamma(\alpha-2)}{\Gamma(2\alpha-2)} \dfrac{1}{r^2} & (\alpha > 2) \end{cases}, \qquad (14)$$

which proves Eq. (8), as promised.

The scaling relation in Eq. (8) can be used for judging if a given sequence is memoryless. Its violation implies memoryfulness. The full expression in Eq. (12) allows one to quantify how a stationary sequence generating the Feller-Pareto distribution is memoryless.

Closing this section, we make a comment on Eq. (6). As mentioned just after Eq. (10), this choice with the manifest invariance in Eq. (5) makes $P(r)$ in Eqs. (9) and (12) independent of the constant $c$. (This may hold also for any function of $x'/x$, in general). It formally becomes "return" in mathematical finance if it is subtracted by unity, provided that $x_k$ in this case should be regarded as the $k$-th price of a financial asset (Mantegna and Stanley 2000). "Price change" $x_{k+m} - x_k$, which then appears in the numerator of return, is also used in that area. It is noted that the variable of price changes is analogous to the velocity difference in physics of turbulence, where the shift is not temporal but spatial (Lewis and Swinney 1999; and the references therein). The variable directly related to that of price change has been applied to seismicity and its modeling (Caruso *et al*. 2007). Although our purpose here is, as emphasized, to characterize the memory and therefore is different from the previous discussions, still we wish to mention that the present work can be viewed as an attempt in the ongoing investigation of understanding similarities between turbulence and financial markets (Mantegna and Stanley 1996) as well as seismicity and brain activity, or scale-invariant systems/phenomena, more generically.

## Applications to earthquake energy and acceleration of ground motion

Now, let us apply this method to examining memories in seismological phenomena.

Firstly, we discuss the sequence of the energy released by each earthquake. The label of the sequence is the event time, that is, $x_k$ stands for the energy of the *k*-th earthquake in the sequence. In Fig. 1, we present the data taken in Japan, which shows the relation between the frequency of the events and the released energies. The time interval is between 00:02:29.62 on 3 June, 2002 and 23:45:9.57 on 31 December, 2007. The geographical region covered is: 17.956N-49.305N latitude, 120.119E-156.047E longitude and depth 0-681 km. The events with magnitude *M* larger than 0.4 are taken (with the definition $\log E = 11.8 + 1.5M$) and the maximum value of magnitude contained is 8.2. The empties (i.e. no records of the values of magnitude) are removed. Correspondingly, the total number of the events included is $\mathcal{N} = 557258$ (recall the notation in the first paragraph in the second section). There, one sees that the frequency well obeys the Feller-Pareto distribution in Eq. (3) and the Gutenberg-Richter law as a pure power law also holds, asymptotically. In Table 1, stationarity (i.e. coincidence of the distributions obtained from the ordered subsequence and its shifted one) of the empirical distributions modeled by the Feller-Pareto form is seen to be valid well for the values of the shift index: $m = 1, 10, 100$, where $N$ is taken to be the total number of the events $\mathcal{N}$ contained in the dataset. Then, in Fig. 2, the results for $P(r)$ calculated from the data and the distribution in Eq. (12), which holds in the memoryless case, are compared. No significant deviations from the memoryless case can be found. Also, the relation in Eq. (8) for the memoryless case holds well since the data values of the exponent are $\alpha = 1.46 \pm 0.01$ for $m = 1, 10, 100$ in consistent with the model value $\alpha = 1.46$, showing that the relation in Eq. (8) to be satisfied in the memoryless case holds well. It is also mentioned that this result turns out to remain unchanged if the value of threshold on magnitude is increased to 2, for example.

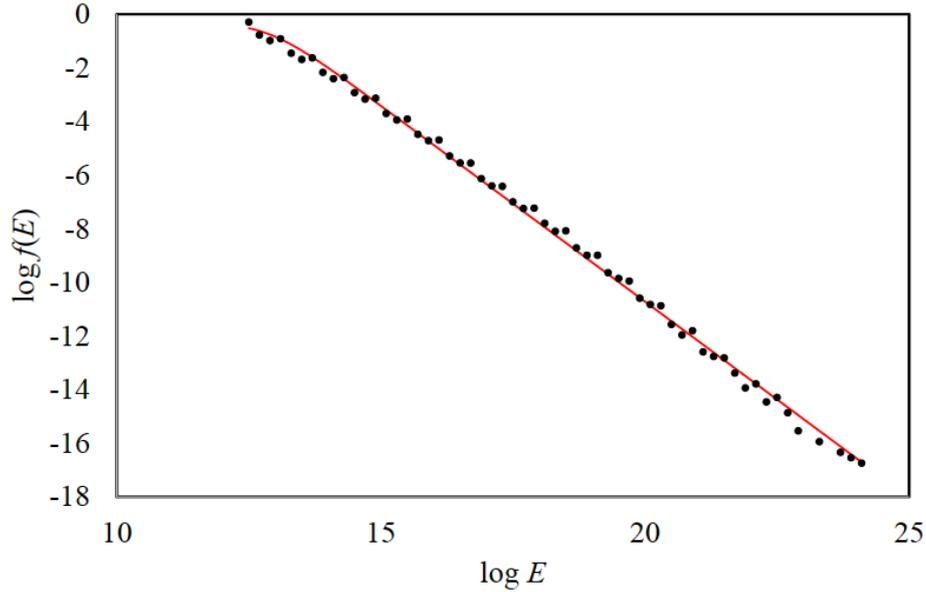

**Fig. 1** The log-log plot of the normalized distribution (i.e. frequency) $f(E)$ [1/J] of earthquakes with the released energies $E$ [J]. The dotted curve and red solid curve respectively describe the real data and the model based on the Feller-Pareto distribution in Eq. (3) with $x = E$, $\alpha = 1.46$, $c = 7.21 \times 10^{12}$ J. The histogram for the real data is made in such a way that the bin size gives five points in each single order of magnitude. The straight line for magnitude larger than 1.6 demonstrates the Gutenberg-Richter law with the $b$-value: $b \approx 0.97$.

**Table 1** Both of the distributions, $f(E)$ [1/J] and its shifted one $f(E')$ [1/J], of earthquakes with respect to their released energies obtained from the ordered subsequences of the whole data set employed in Fig. 1 are well described by the Feller-Pareto form in Eq. (3), which is fully characterized by dimensionless $\alpha$ and $c$ [J]. Here, the values of $\alpha$ and $c$ are presented for the shift index: $m = 1, 10, 100$. Stationarity (i.e. the condition that the distributions of each ordered subsequence and its shifted one coincide each other) is seen to be valid well.

|  |  | $f(E)$ | $f(E')$ |
|---|---|---|---|
| $m=1$: | $\alpha$; | $1.46 \pm 0.01$ | $1.46 \pm 0.01$ |
|  | $c$; | $(7.21 \pm 0.03) \times 10^{12}$ | $(7.21 \pm 0.03) \times 10^{12}$ |
| $m=10$: | $\alpha$; | $1.46 \pm 0.01$ | $1.46 \pm 0.01$ |
|  | $c$; | $(7.21 \pm 0.03) \times 10^{12}$ | $(7.21 \pm 0.03) \times 10^{12}$ |
| $m=100$: | $\alpha$; | $1.46 \pm 0.01$ | $1.46 \pm 0.01$ |
|  | $c$; | $(7.21 \pm 0.03) \times 10^{12}$ | $(7.21 \pm 0.03) \times 10^{12}$ |

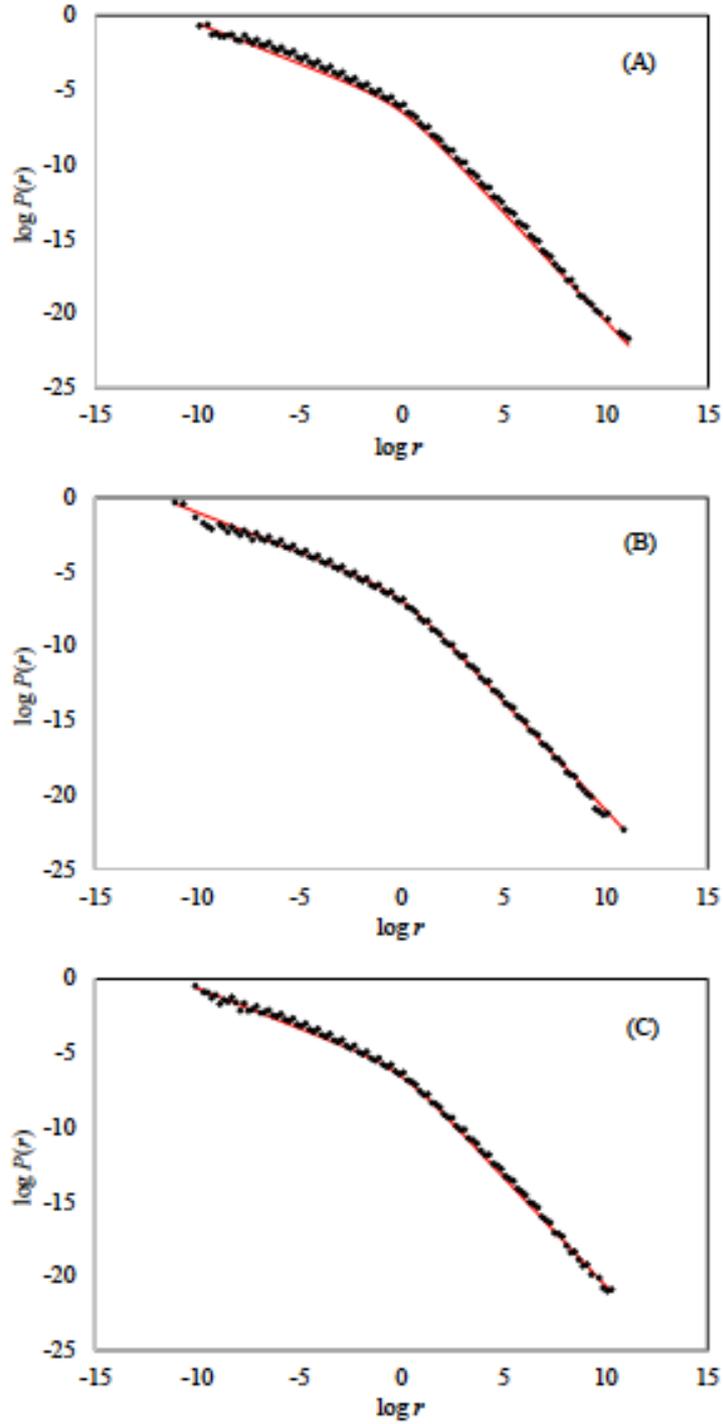

**Fig. 2** The log-log plots of dimensionless $P(r)$ for (A) $m=1$, (B) $m=10$, (C) $m=100$. The dotted curves are the ones calculated by counting of $E'/E(=r)$ from the data, whereas the solid red curves are based on the model in Eq. (12). The model values of $\alpha$ and $c$ employed here are $\alpha=1.46$ and $c=7.21\times10^{12}$ J for (A)-(C), respectively (see Table 1). The histograms of the real data are made in such a way that the bin size gives five points in each single order of magnitude.

Secondly, we discuss the acceleration of ground motion due to a strong earthquake and apply our method to examine its memory property. The label $n$ of the ordered sequence $\{x_n\}_{n=1,2,...,\mathcal{N}}$ is the sampling time, and $x_n$ is the absolute value $|a|$ of the $n$-th sampled value of acceleration of ground motion. In Fig. 3, we present the data taken for the Fukushima-ken Oki earthquake with magnitude 7.4 occurred at 23:36:00 on 16 March, 2022 (37.697N latitude, 141.622E longitude, 57 km in depth), which describes the statistics of the absolute values of the acceleration in the north-south direction sampled at the Iitate station (code FKS004, 37.6798N latitude, 140.7342E longitude, station height 488 m) with the sampling frequency 100 Hz. The data sampled during 23:36.31 and 23:40:09 on 16 March, 2022 is considered, and the values larger than $441.218 \times 10^{-5}$ Gal ($1\,\text{Gal} = 1\,\text{cm/s}^2$) are employed, here. The maximum value is $519.916$ Gal. The total number of the samples is $\mathcal{N} = 21915$ (see the notation given in the second section). Fig. 3 shows how the absolute value of the acceleration well obeys the Feller-Pareto distribution in Eq. (3). In Table 2, stationarity (i.e. coincidence of the distributions obtained from the ordered subsequence and its shifted one) of the empirical distributions modeled by the Feller-Pareto form is seen to be also valid well for the values of the shift index: $m = 1, 10, 100$, where $N$ is taken to be equal to $\mathcal{N}$. Then, in Fig. 4, the results for $P(r)$ are presented. There, the significant deviations from the memoryless case in Eq. (12) are observed. In particular, regarding the exponent, the model values are $\alpha = 1.59$, whereas the data values are $\alpha = 2.23 \pm 0.07$ for $m = 1$, $\alpha = 2.23 \pm 0.03$ for $m = 10$, $\alpha = 2.22 \pm 0.07$ for $m = 100$, showing that the relation in Eq. (8) is violated. Therefore, the sequence of the acceleration is memoryful.

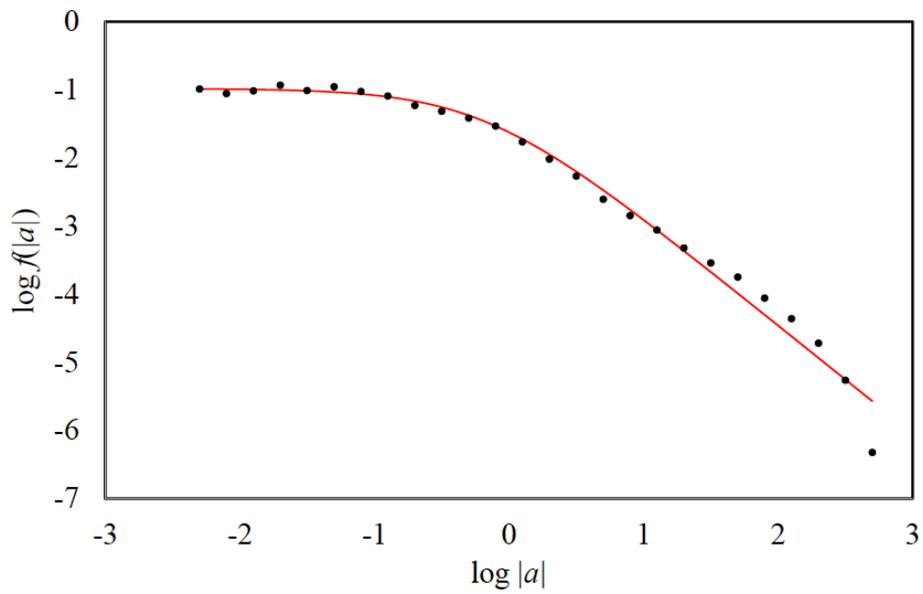

**Fig. 3** The log-log plot of the normalized distribution $f(|a|)$ [1/Gal] with respect to the absolute value of the acceleration $|a|$ [Gal]. The dotted curve and red solid curve respectively describe the real data and the model based on the Feller-Pareto distribution in Eq. (3) with $x=|a|$, $\alpha=1.59$, $c=0.654$ Gal. The histogram for the real data is made in such a way that the bin size gives five points in each single order of magnitude.

**Table 2** Both of the distributions, $f(|a|)$ [1/Gal] and $f(|a'|)$ [1/Gal] ($1\,\text{Gal} = 1\,\text{cm/s}^2$), of the acceleration obtained from the data employed in Fig. 3 are well described by the Feller-Pareto form in Eq. (3), which is fully characterized by dimensionless $\alpha$ and $c$ [Gal]. Here, the values of $\alpha$ and $c$ are presented for the ones of the shift index: $m = 1, 10, 100$. Stationarity (i.e. the condition that the distributions of each ordered subsequence and its shifted one coincide each other) is seen to be valid well.

|  |  | $f(|a|)$ | $f(|a'|)$ |
|---|---|---|---|
| $m = 1$: | $\alpha$; | $1.59 \pm 0.01$ | $1.59 \pm 0.01$ |
|  | $c$; | $(6.54 \pm 0.03) \times 10^{-1}$ | $(6.54 \pm 0.03) \times 10^{-1}$ |
| $m = 10$: | $\alpha$; | $1.59 \pm 0.01$ | $1.59 \pm 0.01$ |
|  | $c$; | $(6.54 \pm 0.03) \times 10^{-1}$ | $(6.55 \pm 0.03) \times 10^{-1}$ |
| $m = 100$: | $\alpha$; | $1.59 \pm 0.01$ | $1.59 \pm 0.01$ |
|  | $c$; | $(6.58 \pm 0.09) \times 10^{-1}$ | $(6.66 \pm 0.09) \times 10^{-1}$ |

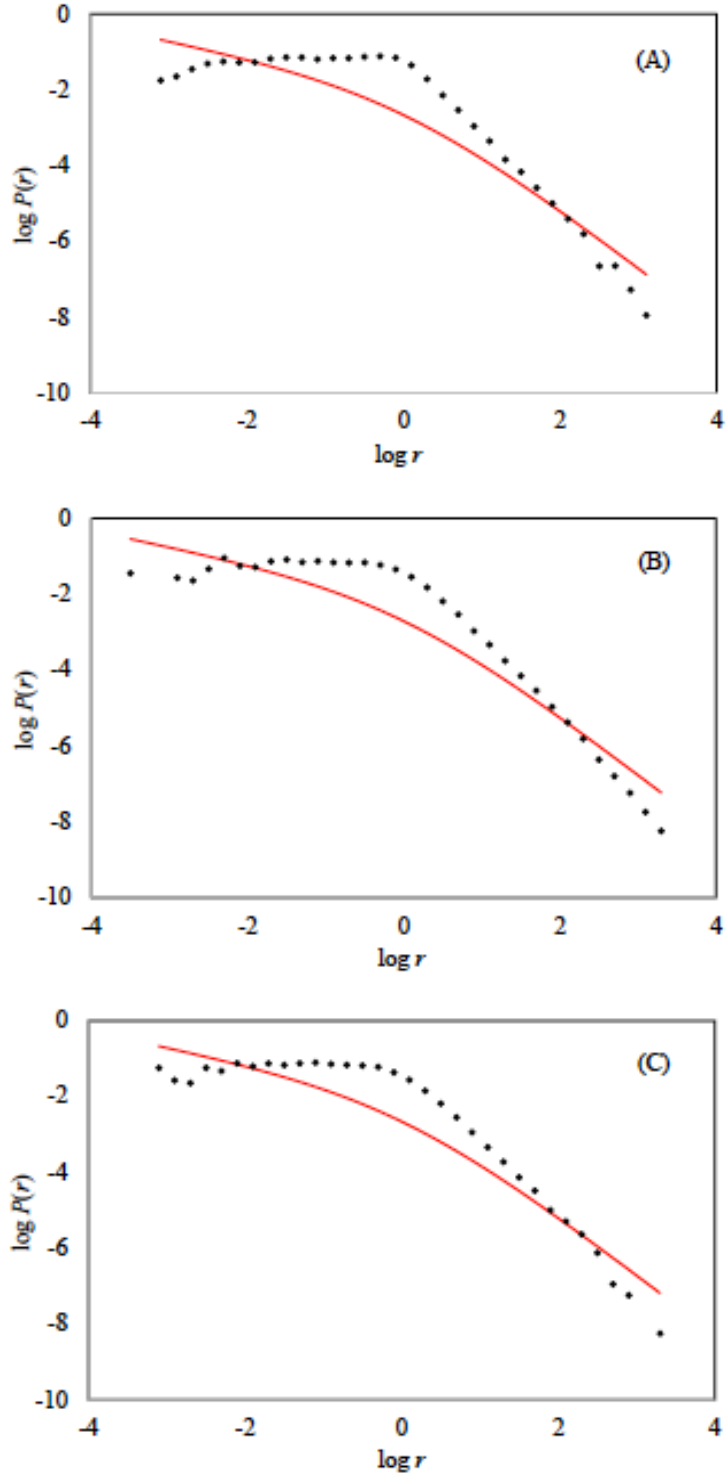

**Fig. 4** The log-log plots of dimensionless $P(r)$ for (A) $m=1$, (B) $m=10$, (C) $m=100$. The dotted curves are the ones calculated by counting of $|a'/a|(=r)$ from the data, whereas the solid red curves are based on the model in Eq. (12). The model values of $\alpha$ and $c$ employed here are $\alpha=1.59$ for (A)-(C), and $c=6.54\times10^{-1}$ Gal for (A) and (B), $c=6.62\times10^{-1}$ Gal for (C), respectively (see Table 2). The histograms

of the real data are made in such a way that the bin size gives five points in each single order of magnitude.

## Concluding remarks

We have formulated a novel theoretical method for characterizing memory of a given ordered sequence generating the stationary distribution of the Feller-Pareto form. In particular, we have derived a simple scaling relation to be satisfied by any memoryless process generating the Feller-Pareto distribution.

In the present work, we have mainly considered the case when the distribution generated by the sequences is of the Feller-Pareto form since such a statistical law is often relevant not only to geophysics but also to other diverse fields in science. However, clearly it is straightforward to generalize our theoretical framework to other distributions. What to be kept is the memoryless condition in Eq. (2), and Eq. (6) should appropriately be modified in accordance with the property of each distribution.

We have applied our method to the sequences of the energies released by earthquakes and the sampled accelerations of ground motion due to a single strong earhtquake. We have shown in a peculiar way that the sequence of the released energies is quite memoryless, whereas the acceleration of ground motion yields a memoryful process. These facts give information on the unknown dynamics governing the processes. The dynamics generating the sequence of the released energy should be strongly chaotic, whereas the acceleration may be relevant to a dynamical system near the onset of chaos where the maximum Lyapunov exponent is vanishingly small and the system remembers its initial condition for a very long duration of time without mixing (Mori *et al.* 1989).

We wish to point out that the results presented here are actually related to the concept of time. In the case of the released energies in seismicity, it is the event time that labels the ordered sequence of earthquakes. On the other hand, time employed in the analysis of the acceleration is the sampling time. And, clearly the scales of these are largely different in terms of the conventional time (i.e. years for the released energies and seconds for the acceleration), and therefore they should not quantitatively be compared with each other, directly. Furthermore, it is noted that the event time is specific to the system/phenomenon, but the sampling time is defined by the measurement condition.

Finally, we make some comments on the scaling relation of $\beta = 2$ ($\alpha > 2$) in Eq. (8) coming from Eq. (14). This relation holds also for distributions different from Eq. (3). For example, consider the exponential and Gaussian distributions in the half space ($x \geq 0$), $f_e(x) = (1/c)\exp(-x/c)$ and $f_G(x) = \sqrt{2/(\pi c^2)}\exp[-x^2/(2c^2)]$. They give rise to $P_e(r) = 1/(1+r)^2$ and $P_G(r) = 2/[\pi(1+r^2)]$, respectively. Both of these asymptotically behave as $P(r) \sim 1/r^2$. This implies that $P(r)$ itself shows how the Feller-Pareto distribution with the exponent $\alpha > 2$ decays "rapidly enough". In fact, the first moment is finite for such an exponent. This point is also natural in view of the generalized central-limit theorem in the half space: multiple convolution of the Feller-Pareto distribution converges to the stable Lévy power-law distribution if and only if $(0<)\alpha<2$ provided that the underlying random variables are independent and identically distributed (Abe and Rajagopal 2000). It is noted that both $P_e(r)$ and $P_G(r)$ are independent of the scale $c$ as Eq. (12) is. This is simply seen as follows. Let us denote Eq. (3), $f_e(x)$ and $f_G(x)$ mentioned above collectively by $f(x;c)$. Clearly, $f(x;c)$ possesses the scaling property: $f(x;c)=(1/c)\tilde{f}(x/c)$, where $\tilde{f}(s)$ stands for the scaling function, which is $\tilde{f}(s)=(\alpha-1)/(1+s)^\alpha$ for Eq. (3), $e^{-s}$ for $f_e(x)$ and $\sqrt{2/\pi}\exp(-s^2/2)$ for $f_G(x)$, respectively. Therefore, the symmetry in Eq. (5) is seen to make $P(r)$ independent of $c$ for any distribution satisfying the scaling property mentioned above.

**Author contributions**   Both of the authors have equally contributed to this work.

**Funding**   S.A. is supported in part by the program of Fujian Province, China and the program of the Kazan Federal University Strategic Academic Leadership Program (PRIORITY-2030). N.S. is indebted to a Grant-in-Aid for Start-up Research of College of Science and Technology, Nihon University.

**Availability of data and material**   The two datasets employed in this work are freely available:

  released energy;  https://hinetwww11.bosai.go.jp/auth/JMA/?LANG=en

  acceleration;  https://www.kyoshin.bosai.go.jp/kyoshin/quake/index_en.html

These are provided by the National Research Institute for Earth Science and Disaster Resilience (NIED). The first catalog is produced by the Japan Meteorological Agency

## Declarations